\documentstyle[12pt,fullpage]{article}

\begin{document}
\begin{flushright}
IJS-TP-99/01\\
TECHNION-PH-99-02\\
hep-ph/9901252\\
\end{flushright}

\vspace{0.3cm}

\begin{center}
{\Large \bf FCNC transitions $c\to u\gamma$ and $s\to d\gamma$ }

\vspace{0.3cm}

{\Large \bf in $B_c\to B_u^*\gamma$ and $B_s\to B_d^*\gamma$ decays }

\vspace{1cm}

{S. Fajfer$^{a,b}$ and S. Prelov\v sek$^{a}$}

\vspace{0.15cm}

{\it a) J. Stefan Institute, Jamova 39, 1001 Ljubljana, Slovenia}

{\it b) Department of Physics, University of Ljubljana, 1000 Ljubljana, Slovenia} 

\vspace{0.5cm}

{P. Singer}

\vspace{0.15cm}

{\it  Department of Physics, Technion - Israel Institute  of Technology, 
Haifa 32000, Israel}
\end{center}

\vspace{0.8cm}

\centerline{\large \bf ABSTRACT}

\vspace{0.3cm}

We propose the $B_c\to B_u^*\gamma$ decay as the most suitable  probe for the flavour changing neutral transition $c\to u\gamma$.  We estimate the short and long distance contributions to this decay within the standard model and we find them to be comparable; this is in contrast to radiative decays of $D$ mesons, that are completely dominated by the long distance contributions. Since the $c\to u\gamma$ transition is very sensitive to the physics beyond the standard model, the standard model prediction $Br(B_c\to B_u^*\gamma)\sim 10^{-8}$ obtained  here opens a new window for future experiments.    
The detection of $B_c\to B_u^*\gamma$ decay at branching ratio well above $10^{-8}$ would signal new physics. In addition we study the $s\to d\gamma$ transition in $B_s\to B_d^*\gamma$ decay and we find it to be dominated by the long distance contribution. We use the Isgur-Scora-Grinstein-Wise (ISGW) constituent quark model for the calculation of these decays. 

\vspace{1.5cm}

\centerline{\bf 1. INTRODUCTION}

\vspace{0.3cm}

Flavour changing neutral current (FCNC) transitions occur in the standard model only at the loop level. Hence, they are very rare in the standard model and they present a suitable probe for new physics. The FCNC transitions in the down-quark sector are relatively frequent due to the large mass of the top quark  running in the loop and the transition $b\to s$  has indeed been observed \cite{CLEO}. The FCNC transitions in the up-quark sector are especially rare in the standard model due to the small masses of the intermediate down-like quarks that run in the loop. For these transitions, the standard model represents a small background for the possible contributions arising from some new physics. At present, only upper experimental limits on the FCNC transitions in the up-quark sector are available \cite{PDG}.

In the present paper we study the transition $c\to u\gamma$, which is the most probable FCNC transition in the up-quark sector within the standard model. To probe the $c\to u\gamma$ transition we propose the radiative beauty-conserving decay $B_c\to B_u^*\gamma$; the  $B_c$ meson has been detected  recently at Fermilab \cite{CDF}. We estimate the short distance (SD) and long distance (LD) contributions to $B_c\to B_u^*\gamma$ decay within the standard model and find them to be comparable, which allows us in principle to probe $c\to u\gamma$ transition in this decay. This is in contrast to previously discussed $D$ meson decays, where the dynamics is completely dominated by the LD contributions [4-7] and it is impossible to extract the short distance  $c\to u\gamma$ contribution from the experiment.

In addition we study $s\to d\gamma$ transition in $B_s\to B_d^*\gamma$ decay. The dynamics of $B_s\to B_d^*\gamma$ is very similar to the dynamics of $B_c\to B_u^*\gamma$ due to the spectator $\bar b$ in both decays. In contrast to $B_c\to B_u^*\gamma$, the decay $B_s\to B_d^*\gamma$ is found to be dominated by LD contributions and consequently the signal of new physics is not expected in this decay.

In Sec. 2 the $B_c\to B_u^*\gamma$ decay is studied: we define SD and LD contributions, present the model and the results. The same reasoning is applied to $B_s\to B_d^*\gamma$ decay is Sec. 3. We conclude with a summary in Sec. 4. \\

\centerline{\bf 2. $B_c\to B_u^*\gamma$ DECAY AND $c\to u\gamma$ TRANSITION}

\vspace{0.3cm}

\centerline{\bf 2a. The short distance contribution}

\vspace{0.3cm}

The SD contribution in $B_c\to B_u^*\gamma$ decay is driven by FCNC $c\to u\gamma$ transition and $\bar b$  is a spectator. The $c\to u\gamma$ transition is strongly GIM suppressed at one-loop, QCD logarithms enhance the amplitude by two orders of magnitude \cite{GHMW}, while the complete 2-loop QCD corrections further increase the amplitude by two orders of magnitude \cite{GHMW}. The Lagrangian that induces the $c\to u\gamma$ transition is given by
\begin{equation}
{\cal L}_{SD}^{c\to u\gamma}=-{G_F\over \sqrt{2}}{e\over 4\pi^2}V_{cs}V_{us}^*~c_7^{c\to u\gamma}(\mu)~\bar u\sigma^{\mu\nu}[m_c{1+\gamma_5\over 2}+m_u{1-\gamma_5\over 2}]c~F_{\mu\nu}~.
\end{equation}
The appropriate scale for $c_7^{c\to u\gamma}(\mu)$ in $B_c\to B_u^*\gamma$ decay is $\mu=m_c$ (and not $\mu=m_b$), since $\bar b$ is merely a spectator in the SD process. The 2-loop QCD calculation was performed in  \cite{GHMW}, giving $c^{c\to u\gamma}_{7}(m_c)=-0.0068-0.020i$. 

The corresponding amplitude for $B_c\to B_u^*\gamma(q,\epsilon)$ decay is proportional to $\epsilon_{\mu}^*q_{\nu}\langle B_u^*|\bar u\sigma^{\mu\nu}(1\pm\gamma_5)c|B_c\rangle$ taken at $q^2=0$, which can be expressed in terms of the form factors $F_1(0)$ and $F_2(0)$ \cite{Soares}:
\begin{eqnarray}
\label{form1}
\epsilon_{\mu}^*\langle B_u^*(p^{\prime},\epsilon^{\prime})|\bar ui\sigma^{\mu\nu}q_{\nu}c|B_c(p)\rangle_{q^2=0}&=&i\epsilon^{\mu\alpha\beta\gamma}\epsilon_{\mu}^*\epsilon_{\alpha}^{*\prime}p_{\beta}^{\prime}p_{\gamma}F_1(0)~,\nonumber\\
\epsilon_{\mu}^*\langle B_u^*(p^{\prime},\epsilon^{\prime})|\bar ui\sigma^{\mu\nu}q_{\nu}\gamma_5c|B_c(p)\rangle_{q^2=0}&=&[(m_{B_c}^2-m_{B_u^*}^2)\epsilon^*\cdot\epsilon^{*\prime}-2(\epsilon^{*\prime}\cdot q)(p\cdot \epsilon^{*})]F_2(0).
\end{eqnarray}
 The form factors defined in Eqs. (\ref{form1})  will be calculated using the ISGW model \cite{ISGW} later on in this section. \\

\centerline{\bf 2b. The long distance contributions}

\vspace{0.3cm}

The long distance contributions will be calculated by using the nonleptonic weak Lagrangian, which on the quark level can be written as \cite{FS} 
\begin{equation}
\label{LD}
{\cal L}^{eff}(\Delta c=1)=-{G_F\over \sqrt{2}}V_{uq_i}V^*_{cq_j}[a_1(\bar u q_i)^{\mu}_{V-A}(\bar q_jc)_{V-A,\mu}+a_2(\bar u c)^{\mu}_{V-A}(\bar q_jq_i)_{V-A,\mu}]~,
\end{equation}
where $(\bar \psi_1\psi_2)^{\mu}_{V-A}=\bar\psi_1\gamma^{\mu}(1-\gamma_5)\psi_2$, $q_i$, $q_j$ are the down quarks $d$, $s$, $b$ and $a_1$, $a_2$ include the QCD corrections \cite{BSW}.

Quite generally, the LD contributions to $B_c\to B_u^*\gamma$ decay can be separated into two classes related to the two terms of  (\ref{LD}), as performed previously \cite{BGHP} for $D\to V\gamma$ decays. The class (I), called also the VMD contribution, is related to the $a_2$ term (\ref{LD}) and  corresponds to the processes $c\to u\bar q_iq_i$ followed by $\bar q_iq_i\to \gamma$, while $\bar b$ is the spectator in $B_c\to B_u^*\gamma$ decay. At the hadron level the $\bar q_iq_i\to \gamma$ transition is expressed using the vector meson dominance (VMD) and the corresponding diagram is depicted in Fig. (1a). The class (II), called also the {\it pole} contribution, is related to the $a_1$ term (\ref{LD}) and  corresponds to the process $c\bar b\to u\bar b$ with the photon attached to incoming or outgoing quark lines. Selecting the lowest contributing states, the pole contributions are depicted in Fig. (1b). 

We turn now to the estimation of these two classes of contributions and we start with the VMD contribution (class (I)) represented by Fig. (1a). The underlying quark processes are $c\to u\bar ss(\bar dd)$ with $\bar ss,~\bar dd$ hadronizing into vector mesons $\phi$, $\rho$, $\omega$ which then turn to a photon, while $\bar b$ remains a spectator. We neglect the contribution of $\bar bb\to \gamma$ in view of the large mass of $\Upsilon$. The relevant part of the Lagrangian, after using the relations among CKM matrix elements, is
\begin{eqnarray}
\label{L1}
{\cal L}_{(I)}^{eff}=-{G_F\over \sqrt{2}}a_2(\mu)V_{cs}V_{us}^*~\bar u\gamma^{\mu}(1-\gamma_5)c~[\bar s\gamma_{\mu}(1-\gamma_5)s-\bar d\gamma_{\mu}(1-\gamma_5)d]~.
\end{eqnarray}
The appropriate scale for $a_2(\mu)$ in $B_c\to B_u^*\gamma$ decay is $\mu=m_c$, since $\bar b$ is again merely a spectator in VMD contribution. 
Thus, we may use $a_2(m_c)=-0.5$, as obtained in the successful phenomenological fit to $D$ meson decays \cite{BSW}. Defining $\langle V(q,\epsilon)|V_{\mu}|0\rangle=g_V(q^2)\epsilon_{\mu}^*$ and using the factorization approximation, the effective Lagrangian that induces the VMD contribution is given by 
\begin{eqnarray}
\label{VMD}
{\cal L}_{VMD}^{c\to u\gamma(\epsilon)}=-{G_Fe\over \sqrt{2}}a_2(m_c)V_{cs}V_{us}^*C_{VMD}^{'}~\bar u\gamma^{\mu}(1-\gamma_5)c~\epsilon_{\mu}^*~,
\end{eqnarray}  
were
\begin{equation} 
\label{cvmd}
C_{VMD}^{'}={g_{\rho}^2(0)\over  2m_{\rho}^2}-{g_{\omega}^2(0)\over  6m_{\omega}^2}-{g_{\phi}^2(0)\over  3m_{\phi}^2}=(-1.2\pm 1.2)\cdot 10^{-3}~GeV^2~
\end{equation}
is obtained by assuming  $g_{V}(m_{V})=g_{V}(0)$, with the mean value and the error in (\ref{cvmd}) calculated from the experimental data on $\Gamma(V\to e^+e^-)$ \cite{PDG}. Note here the remarkable GIM cancellation carried over to the hadronic level. 

Lagrangian (\ref{VMD}) implies that the  VMD amplitude for $B_c\to B_u^*\gamma(q,\epsilon)$ is proportional to  $\epsilon_{\mu}^*\langle B_u^*|\bar u\gamma^{\mu}(1-\gamma_5)c|B_c\rangle$ taken at $q^2=0$. For the hadronic matrix elements, one defines appropriate form factors for the vector and axial transitions as follows \cite{Soares}:
\begin{eqnarray} 
\label{form2}
\langle B_u^*(p^{\prime},\epsilon^{\prime})\!\!\!\!&|&\!\!\!\!\!\bar u\gamma^{\mu}(1-\gamma_5)c|B_c(p)\rangle=-{2i\over m_{B_c}+m_{B_u^*}}\epsilon^{\mu\alpha\beta\gamma}\epsilon_{\alpha}^{*\prime}p_{\beta}^{\prime}p_{\gamma}V(q^2)+(m_{B_c}+m_{B_u^*})\epsilon^{\mu *\prime}A_1(q^2)\nonumber\\
&\!\!\!-\!\!\!\!&{\epsilon^{*\prime}\cdot q\over m_{B_c}+m_{B_u^*}}(p+p^{\prime})^{\mu}A_2(q^2)
-2m_{B_u^*}{\epsilon^{*\prime}\cdot q\over q^2}q^{\mu}[A_3(q^2)-A_0(q^2)]~.
\end{eqnarray}
The requirements of the finite matrix elements at $q^2=0$ \cite{BSW} and of gauge invariance lead to the relations among the various form factors \cite{FPS1}, which imply $A_0(0)=A_3(0)=0$ and $A_2(0)=[(m_{B_c}+m_{B_u^*})/(m_{B_c}-m_{B_u^*})]A_1(0)$. The same relations are obtained by using the prescription that the photon couples only to the transverse polarization of the current \cite{FPS1,GP}. Accordingly, the VMD amplitude will  be expressed in terms of two form factors only, $V(0)$ and $A_1(0)$. 

At this point, we remark that the form factors $F_1$, $F_2$, $V$ and $A_1$, needed for the SD and VMD amplitudes cannot be safely related using the Isgur-Wise relations \cite{IW}, since the masses of $b$ and $c$ quarks composing $B_c$ meson do not permit the $\bar b$ quark to be at rest. Therefore we shall determine the corresponding form factors at $q^2=0$ independently, using the ISGW model \cite{ISGW}.

\vspace{0.2cm}

We now turn to the discussion of the LD contributions of class (II), the {\it pole} contribution, where the quark process $c \bar b \to u\bar b$ is driven by
\begin{equation}
\label{pole}
{\cal L}_{(II)}^{eff}=-{G_F\over \sqrt{2}}a_1(\mu)V_{cb}V_{ub}^*~\bar u\gamma^{\mu}(1-\gamma_5)b~\bar b\gamma_{\mu}(1-\gamma_5)c~
\end{equation}
and the photon line is attached to any of four quark lines.  In terms of hadronic degrees of freedom this diagram is given in Fig. (1b), where the white box represents the action of the Lagrangian (\ref{pole}) (we have neglected the contribution of the scalar and axial poles).  
Considering the scale for $a_1(\mu)$ in $c\bar b\to u\bar b$, it is difficult to decide between $\mu=m_c$ or $\mu=m_b$, since $\bar b$ is not spectator in the pole contribution. As the difference between  $a_1(m_c)=1.2$ and $a_1(m_b)=1.1$ \cite{BSW} and is not essential, we take $a_1(m_b)=1.1$. 
Note that the pole contribution is relatively small due to the factor $V_{cb}V_{ub}^*$ in (\ref{pole}). In $D$ meson decays, the corresponding factor $V_{cs}V_{us}^*$ is much bigger, which makes the pole contribution dominant  over the SD and VMD ones \cite{BGHP,FS,FPS1}. The different CKM matrix elements in the pole contributions of $B_c$ and $D$ decays is a major factor in establishing 
the $B_c\to B_u^*\gamma$ decay as more suitable  for  the investigation of $c\to u\gamma$ than the $D$ decays.

To evaluate the amplitude for the pole diagrams given in Fig. (1b) we define    \begin{eqnarray}
\label{pole1}
\langle 0|A_{\mu}|P\rangle &=&f_Pp_{\mu}\\
\langle V|V_{\mu}|0\rangle &=&g_V\epsilon_{\mu}^*\nonumber\\
{\cal A}(P(p)\to V(p^{\prime},\epsilon^{\prime})\gamma(\epsilon ))&=&\mu_Pe\epsilon^{\mu\nu\alpha\beta}\epsilon_{\mu}^*\epsilon_{\nu}^{*\prime}p_{\alpha}p_{\beta}^{\prime}~,\nonumber
\end{eqnarray}
where $\mu_{B_c}$, $\mu_{B_u}$, $f_{B_c}$, $f_{B_u}$, $g_{B_c^*}$ and $g_{B_u^*}$ will be determined using ISGW model.\\

\centerline{\bf 2c. The amplitude}

\vspace{0.3cm}

Using the above Lagrangians and form factor decomposition of Eqs. (\ref{form1}), (\ref{form2}), (\ref{pole1}), the final amplitude for $B_c\to B_u^*\gamma$ containing SD and LD contributions can be expressed as
\begin{equation}
\label{amp}
A(B_c(p)\to B_u^*(p^{\prime},\epsilon^{\prime})\gamma(q,\epsilon ))=i\epsilon_{\mu}^{*\prime}\epsilon_{\nu}^*[A_{PV}(p^{\mu}p^{\nu}-g^{\mu\nu}p\cdot q)+iA_{PC}\epsilon^{\mu\nu\alpha\beta}p^{\prime}_{\alpha}p_{\beta}]~,
\end{equation}
where
\begin{eqnarray}
\label{e1}
A_{PV}&=&-{G_F\over \sqrt{2}} e\biggl(V_{cs}V_{ud}^*\biggl[{c_7^{c\to u\gamma}(m_c)\over 2\pi^2}(m_c-m_u)F_2(0)+2a_2(m_c)C_{VMD}^{\prime}{A_1(0)\over m_{B_c}-m_{B_u^*}}\biggr]\biggr)~,\nonumber\\
A_{PC}&=&-{G_F\over \sqrt{2}} e\biggl(V_{cs}V_{ud}^*\biggl[{c_7^{c\to u\gamma}(m_c)\over 4\pi^2}(m_c+m_u)F_1(0)+2a_2(m_c)C_{VMD}^{\prime}{V(0)\over m_{B_c}+m_{B_u^*}}\biggr]\nonumber\\
&+&V_{cb}V_{ub}^*a_1(m_b)\biggl[{\mu_{B_c}g_{B_c^*}g_{B_u^*}\over m_{B_c^*}^2-m_{B_u^*}^2}+{\mu_{B_u}m_{B_c}^2f_{B_c}f_{B_u}\over m_{B_c}^2-m_{B_u}^2}\biggr]
\biggr)~.
\end{eqnarray}
The first term in Eqs. (\ref{e1}) comes from SD contribution, the second term from VMD contribution and the third term from the {\it pole} contribution. The decay width is then given by
\begin{equation}
\label{gamma}
\Gamma={1\over 4\pi}\biggl({m_{B_c}^2-m_{B_u^*}^2\over 2m_{B_c}}\biggr)^3(|A_{PV}|^2+|A_{PC}|^2)~.
\end{equation}

\vspace{0.3cm}

\centerline{\bf 2d. The model}

\vspace{0.3cm}

To account for the nonperturbative dynamics within the mesons we use the nonrelativistic constituent ISGW quark model \cite{ISGW}. This model is considered  to be reliable for a state composed of two heavy quarks, which makes it suitable for treating $B_c$; in addition the velocity of $B_u^*$ in the rest frame of $B_c$ is to a fair approximation nonrelativistic. In the ISGW model the constituent quarks of mass $M$ move under the influence of the effective potential $V(r)=-4\alpha_s/(3r)+c+br$, $c=-0.81~GeV$, $b=0.18~GeV^2$ \cite{ISGW2}. Instead of the accurate solutions of the Schrodinger equation, the variational solutions 
$$\psi(\vec r)=\pi^{-{3\over 4}}\beta^{{3\over 2}}e^{-{\beta^2r^2\over 2}}~~~{\rm or}~~~ \psi(\vec k)=\pi^{-{3\over 4}}\beta^{-{3\over 2}}e^{-{k^2\over 2\beta^2}}~~~~~~{\rm for~~ S~~ state}$$
are used, where $\beta$ is employed as the variational parameter. The meson state composed of constituent quarks $q_1$ and $\bar q_2$ is given by   
\begin{equation}
|M(p)\rangle=\sum_{C,s1,s2}{1\over \sqrt{3}}\sqrt{{2E\over (2\pi)^3}}\int d\vec k\psi(\vec k)\sqrt{{M_1\over E_1}}\sqrt{{M_2\over E_2}}f_{s2,s1}\delta(p-p_1-p_2)b_1^\dagger(\vec p_1,s_1,C)d_2^\dagger(\vec p_2,s_2,\bar C)|0\rangle~,
\end{equation} 
where $\vec k$ is the momentum of the constituents in the meson rest frame, $C$ denotes the colour, while $f_{s2,s1}=(\bar\uparrow\downarrow+\bar\downarrow\uparrow)/\sqrt{2}$ for pseudoscalar and $f_{s2,s1}=(\bar\uparrow\downarrow-\bar\downarrow\uparrow)/\sqrt{2},\bar\uparrow\uparrow,\bar\downarrow\downarrow$ for vector mesons. Using the normalization of the spinors as in \cite{itzykson}, we obtain in the nonrelativistic limit 
\begin{eqnarray}
V(q^2)&=&{m_{B_c}+m_{B_u^*}\over2}F_3(q^2)\biggl[{1\over M_u}-{M_b(M_c-M_u)\beta_{B_c}^2\over M_cM_um_{B_u^*}(\beta_{B_c}^2+\beta_{B_u^*}^2)}\biggr]~,\nonumber\\
A_1(q^2)&=&F_2(q^2)={2m_{B_c}\over m_{B_c}+m_{B_u^*}}F_3(q^2)~,\nonumber\\
F_1(q^2)&=&2F_3(q^2)\biggl[1+(m_{B_c}-m_{B_u^*})\biggl({1\over 2M_u}-{M_b(M_c+M_u)\beta_{B_c}^2\over 2M_cM_um_{B_u^*}(\beta_{B_c}^2+\beta_{B_u^*}^2)}\biggr)\biggr]~,\nonumber\\
\mu_{B_c}&=&\sqrt{{m_{B_c^*}\over m_{B_c}}}\biggl({2\beta_{B_c}\beta_{B_c^*}\over \beta_{B_c}^2+\beta_{B_c^*}^2}\biggr)^{3\over 2}\biggl[{2\over 3M_c}-{1\over 3M_b}\biggr]~,\nonumber\\
f_{B_c}&=&{2\sqrt{3}\beta_{B_c}^{{3\over 2}}\over \pi^{{3\over 4}}\sqrt{m_{B_c}}}~,\nonumber\\
g_{B_c^*}&=&m_{B_c^*}{2\sqrt{3}\beta_{B_c^*}^{{3\over 2}}\over \pi^{{3\over 4}}\sqrt{m_{B_c^*}}}
\end{eqnarray}
and analogously for $\mu_{B_u}$, $f_{B_u}$ and $g_{B_u^*}$. Here 
$$F_3(q^2)=\sqrt{m_{B_u^*}\over m_{B_c}}\biggl({2\beta_{B_c}\beta_{B_u^*}\over \beta_{B_c}^2+\beta_{B_u^*}^2}\biggr)^{3/2}\exp\biggl(-{M_b^2\over 2m_{B_c}m_{B_u^*}} {[(m_{B_c}-m_{B_u^*})^2-q^2]\over \kappa^2(\beta_{B_c}^2+\beta_{B_u^*}^2)}\biggr)~,$$
where $\kappa=0.7$ \cite{ISGW}. The results for $V(q^2)$ and $A_1(q^2)$ reproduce the results of \cite{ISGW}, while $F_1(q^2)$ and $F_2(q^2)$ represent, to our knowledge, the new results within ISGW model. Using parameters $\beta$ \cite{ISGW2} and meson masses given in Table 1  and the constituent quark masses $M_u=0.33~GeV,~~M_c=1.82~GeV~~{\rm and}~~M_b=5.2~GeV$ \cite{ISGW2} we get
$$f_{B_u}=0.18~GeV~,~~g_{B_u^*}=0.86~GeV^2~,~~\mu_{B_u}=1.81~GeV^{-1}$$
$$f_{B_c}=0.51~GeV~,~~g_{B_c^*}=2.41~GeV^2~,~~\mu_{B_c}=0.28~GeV^{-1}~,$$
while the form factors evaluated at $q^2=0$ are given in Table 2. \\

\centerline{\bf 2e. The results for $B_c\to B_u^*\gamma$}

\vspace{0.3cm}

We use the central value of the current quark masses $m_u=0.0035~GeV$, $m_c=1.25~GeV$ from \cite{PDG} and $V_{cb}=0.04$, $V_{ub}=0.0035$. The SD, VMD and {\it pole} contributions to amplitudes $A_{PC}$ and $A_{PV}$ needed to compute the amplitude (\ref{amp}) and the decay rate (\ref{gamma}) are given in Table 3, where the error is due only to the uncertainty in parameter $C_{VMD}^{\prime}$ (\ref{cvmd}). In Table 4 we present the total branching ratio and separately also the SD and LD part of the branching ratios for $B_c\to B_u^*\gamma$ decay, where we have taken $\tau(B_c)=0.46{+0.18\atop -0.16}\pm0.03~ps$ as measured by CDF Collaboration recently \cite{CDF}. Note that SD and LD contributions  give branching ratios of comparable size $\sim 10^{-8}$, which in principle allows to probe the $c\to u\gamma$ transition in $B_c\to B_u^*\gamma$ decay. Experimental detection of $B_c\to B_u^*\gamma$ decay at the branching ratio well above $10^{-8}$ would clearly indicate a signal for new physics.  The measurement of this decay would probe different scenarios of  physics beyond the standard model: the non-minimal supersymmetric model \cite{BGM} and the standard model with four generations \cite{BHLP}, for example, predict  $Br(c\to u\gamma)$ up to $10^{-5}$, which would enhance $Br(B_c\to B_u^*\gamma)$ up to $10^{-6}$.   In $D$ meson decays ($c\bar q\to u\bar q\gamma$), on the other hand, the branching ratios are of order $10^{-6}$ even within the standard model \cite{FS,FPS1,GHMW}: they are driven mainly by the long distance pole contributions, which overshadow the  $c\to u\gamma$ transition (predicted at the branching ratio $\sim 10^{-9}$ in the standard model) and possible signals of new physics.\\

\centerline{\bf 3. $B_s\to B_d^*\gamma$ DECAY AND $s\to d\gamma$ TRANSITION}

\vspace{0.3cm}

The calculation of SD and LD contributions to $B_s\to B_d^*\gamma$ decay is analogous to what was presented in the previous section. Due to the small difference $m_{B_s}-m_{B_d^*}$, the final $B_d^*$ meson is almost at rest in the rest frame of $B_s$. 

The SD Lagrangian is 
\begin{equation}
\label{SD1}
{\cal L}_{SD}^{s\to d\gamma}=-{G_F\over \sqrt{2}}{e\over 4\pi^2}V_{cs}V_{us}^*c_7^{s\to d\gamma}(m_s)~\bar d\sigma^{\mu\nu}[m_s{1+\gamma_5\over 2}+m_d{1-\gamma_5\over 2}]s~F_{\mu\nu}~,
\end{equation}
where $c_7^{s\to d\gamma}(m_s)=-0.23$ \cite{Singer}.

The Lagrangian, from which the long distance VMD contribution is calculated, is given by   
\begin{eqnarray}
{\cal L}_{VMD}^{s\to d\gamma}&=&-{G_F\over \sqrt{2}}a_2(m_s)V_{ud}V_{us}^*~\bar d\gamma^{\mu}(1-\gamma_5)s~[\bar u\gamma_{\mu}(1-\gamma_5)u-\bar c\gamma_{\mu}(1-\gamma_5)c]~,
\end{eqnarray}
were $|a_2(m_s)|\sim 0.5$ \cite{Singer} and 
\begin{equation}
C_{VMD}={g_{\rho}^2(0)\over  2m_{\rho}^2}+{g_{\omega}^2(0)\over  6m_{\omega}^2}-\sum_i{2g_{\psi i}^2(0)\over  3m_{\psi i}^2}~.
\end{equation}
From $\Gamma_{exp}(\Omega^-\to \Xi^-\gamma)<3.7\cdot 10^{-9} ~eV$ the upper limit $|C_{VMD}|<0.01~GeV^2$ has been obtained \cite{Singer}. 

The {\it pole} contribution is absent in $B_s\to B_d^*\gamma$ decay since the decay $s\bar b\to d\bar b$ involves four quarks of the same charge. 

\vspace{0.1cm}

We use the constituent quark masses $M_d=0.33~GeV$, $M_s=0.55~GeV$ \cite{ISGW2}, the central value of the current quark masses $m_d=0.006~GeV$, $m_s=0.115~GeV$ \cite{PDG} and the masses and parameters $\beta$ of the mesons given in Table 1. Using the formulas of the previous section and discarding  the pole amplitudes, the resulting amplitudes $A_{PC,PV}$ and the branching ratios are given in Tables 3 and 4, respectively, where the upper limits are due to $|C_{VMD}|<0.01~GeV^2$. The  upper limit $Br(B_s\to B_d^*\gamma)<1.8\cdot 10^{-7}$ is dominated by the LD contribution. The same conclusion with a smaller upper bound was obtained in \cite{Singer} assuming the simple free quark decay, which is a reasonable assumption for $s\to d\gamma$ in $B_s\to B_d^*\gamma$ decay. However, the calculation of the VMD amplitude in \cite{Singer} was based on the formalism presented in \cite{DHT}, which is not reliable for $s\to d\gamma$ transition.  \\

\centerline{\bf 4. SUMMARY} 

\vspace{0.3cm}

We have studied flavour changing neutral transitions $c\to u\gamma$ and $s\to d\gamma$ in $B_c\to B_u^*\gamma$ and $B_s\to B_d^*\gamma$ decays, respectively. The predicted short and long distance contributions to these decays within the standard model are presented in Table 4. 

We predict $Br(B_s\to B_d^*\gamma)<1.8\cdot 10^{-7}$, which is dominated by the long distance contribution.

The short distance part (driven by $c\to u\gamma$) and the long distance part of the branching ratio for $B_c\to B_u^*\gamma$ decay are found to be of comparable size; they are both of order $10^{-8}$. This makes $B_c\to B_u^*\gamma$ decay the most suitable decay to probe the $c\to u\gamma$ transition. Since $c\to u\gamma$ transition is very sensitive to the physics beyond the standard model, it would be very desirable to compare the standard model prediction of $Br(B_c\to B_u^*\gamma)=(8.5 {+5.8 \atop -2.5})\cdot 10^{-9}$ presented here to the experimental data in the future.    
The detection of $B_c\to B_u^*\gamma$ decay at a branching ratio well above $10^{-8}$ would signal new physics. In comparison to $B_c\to B_u^*\gamma$ decay, the  $D$ meson decays are far less suitable for probing $c\to u\gamma$ transition, since they are almost completely dominated by the long distance effects.

Finally, we wish to stress that $B_c\to B_u^*\gamma$ and $B_s\to B_d^*\gamma$ decays are characterized by a very clear
signature: their detection requires the observation of a $B_u/B_d$ decay in
coincidence with two photons. The $B_c\to B_u^*\gamma$  transition involves the emission of a high energy ($985~ MeV$) and of a low energy ($45~ MeV$) photon in the respective centers of mass of $B_c$, $B_u^*$, while in  the decay 
$B_s\to B_d^*\gamma$ two photons of nearly equal energy ($45 ~MeV$) are emitted.\\

\centerline{\bf ACKNOWLEDGMENTS}

\vspace{0.3cm}

The research of S.P. and S.F. was supported in part by the Ministry of Science of the Republic of Slovenia. 
The research of P.S. was supported in part by the Fund for 
Promotion of Research at the Technion.\\

{\bf Figure caption}

\vspace{0.2cm}

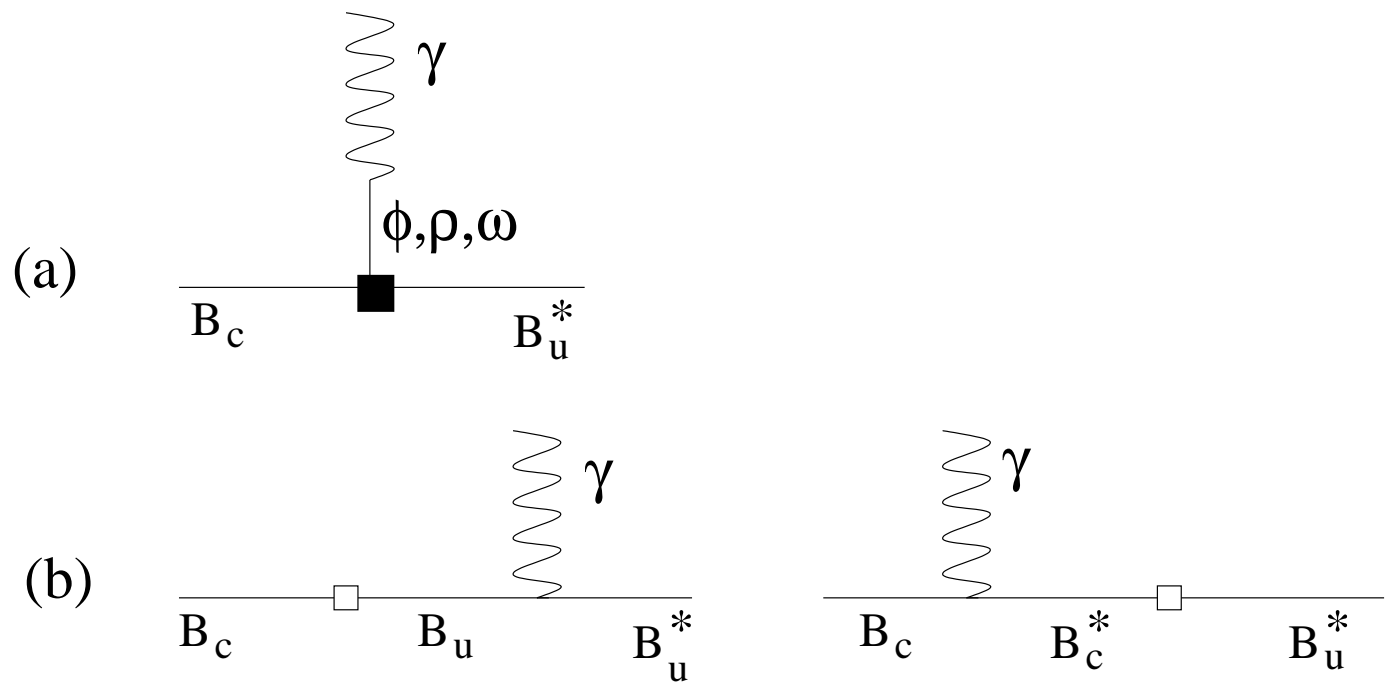
{\bf Fig. 1:} Long distance contributions in $B_c\to B_u^*\gamma$ decay. a) VMD contribution; the black box denotes the action of the Lagrangian (\ref{VMD}). b) {\it pole} contribution; the white box denotes the action of the Lagrangian (\ref{pole}).\\

\begin{table}[h]
\begin{center}
\begin{tabular}{|c||c|c|c|c|c|c|}
\hline
& $B_c$&$B_c^*$&$B_u$&$B_u^*$&$B_s$&$B_d^*$\\
\hline
$m$&6.40 \cite{CDF}&6.42 \cite{ISGW2}&5.28 \cite{PDG}&5.325 \cite{PDG}&5.37 \cite{PDG}&5.325 \cite{PDG}\\
\hline
$\beta$&0.92&0.75&0.43&0.40&0.54&0.40\\
\hline
\end{tabular}
\caption{Parameters $\beta$ (taken from \cite{ISGW2})  and masses of pseudoscalar and vector mesons in GeV.  }
\end{center}
\end{table}

\begin{table}[h]
\begin{center}
\begin{tabular}{|c||c|c|c|c|}
\hline
&$A_1(0)$&$V(0)$&$F_1(0)$&$F_2(0)$\\
\hline
$B_c\to B_u^*$&0.24&1.3&0.48&0.24\\
\hline
$B_s\to B_d^*$&0.90&11&1.8&0.90\\
\hline
\end{tabular}
\caption{Form factors at $q^2=0$ calculated using ISGW model \cite{ISGW}.  }
\end{center}
\end{table}

\begin{table}[h]
\begin{center}
\begin{tabular}{|c||c|c|c||c|c|c|}
\hline
& $A_{PV}^{SD}$ & $A_{PV}^{VMD}$ & $A_{PV}^{pole}$ & $A_{PC}^{SD}$ & $A_{PC}^{VMD}$ & $A_{PC}^{pole}$ \\
\hline
$B_c\to B_u^*\gamma$&$5.7+17~i$&$-14\pm 14$&$0$&$5.7+17~i$&$-7.3\pm 7.3$&$-21$\\
\hline 
$B_s\to B_d^*\gamma$&$62$&$<1.1\cdot 10^{4}$&$0$&$70$&$<5.6\cdot 10^{2}$&$0$\\
\hline
\end{tabular}
\caption{ The amplitudes $A_{PV,PC}$ defined in (\ref{amp}) for SD, VMD and {\it pole} contributions in units of $10^{-11}~GeV^{-1}$ as predicted by ISGW model. The error-bars in $B_c\to B_u^*\gamma$ are due to the uncertainty in $C_{VMD}^{'}=(1.2\pm 1.2)~10^{-3}~GeV^2$ (\ref{cvmd}), while the upper bounds for $B_s\to B_d^*\gamma$ are due to $|C_{VMD}|<0.01~GeV^2$.}
\end{center}
\end{table}

\begin{table}[h]
\begin{center}
\begin{tabular}{|c||c|c|c|}
\hline
& $Br^{SD}$ & $Br^{LD}$ & $Br^{tot}$ \\
\hline
$B_c\to B_u^*\gamma$& $4.7\cdot 10^{-9}$ & $(7.5 {+7.7\atop -4.3})\cdot 10^{-9}$ & $(8.5 {+5.8 \atop -2.5})\cdot 10^{-9}$\\
\hline 
$B_s\to B_d^*\gamma$& $1.4\cdot 10^{-11}$ & $<2.0\cdot 10^{-7}$ & $<2.0\cdot 10^{-7}$\\ 
\hline
\end{tabular}
\caption{ SD, LD and total branching ratios as predicted by ISGW model. The error-bars in $B_c\to B_u^*\gamma$ are due to the uncertainty in $C_{VMD}^{'}=(1.2\pm 1.2)~10^{-3}~GeV^2$ (\ref{cvmd}), while the upper bounds for $B_s\to B_d^*\gamma$ are due to $|C_{VMD}|<0.01~GeV^2$.}
\end{center}
\end{table}

\end{document}